\documentclass [11pt, letterpaper]{article}
\usepackage{fullpage}
\usepackage{amsmath}
\usepackage{amssymb}
\usepackage{graphicx}
\usepackage{mathrsfs}
\usepackage{wrapfig}
\usepackage{boxedminipage}
\usepackage{epsfig}
\usepackage{setspace}
\usepackage{subfigure}
\usepackage{float}
\usepackage{hyperref}
\usepackage{enumerate}
\newcommand{\reef}[1]{(\ref{#1})}
\begin{document}

\begin{flushright}
\phantom{{\tt arXiv:1105.????}}
\end{flushright}

\bigskip
\bigskip
\bigskip

\begin{center} {\Large \bf Thermal Dynamics of  Quarks and Mesons}
  
  \bigskip

{\Large\bf  in }

\bigskip

{\Large\bf     $\mathcal{N}=2^\ast$ Yang--Mills Theory}

\end{center}

\bigskip \bigskip \bigskip \bigskip

\centerline{\bf Tameem Albash, Clifford V. Johnson}

\bigskip
\bigskip
\bigskip

  \centerline{\it Department of Physics and Astronomy }
\centerline{\it University of
Southern California}
\centerline{\it Los Angeles, CA 90089-0484, U.S.A.}

\bigskip

\centerline{\small \tt talbash,  johnson1,  [at] usc.edu}

\bigskip
\bigskip


\begin{abstract} 
\noindent 
We study the dynamics of quenched fundamental matter in
$\mathcal{N}=2^\ast$ supersymmetric large $N_c$ $SU(N_c)$ Yang--Mills
theory, extending our earlier work to finite temperature. We use
probe D7--branes in the holographically dual thermalized
generalization of the $\mathcal{N}=2^\ast$ Pilch--Warner gravitational
background found by Buchel and Liu. Such a system provides an
opportunity to study how key features of the dynamics are affected by
being in a non--conformal setting where there is an intrinsic scale,
set here by the mass, $m_H$, of a hypermultiplet. Such studies are
motivated by connections to experimental studies of the quark--gluon
plasma at RHIC and LHC, where the microscopic theory of the
constituents, QCD, has a scale, $\Lambda_{\rm QCD}$. We show that the
binding energy of mesons in the $\mathcal{N}=2^\ast$ theory is
increased in the presence of the scale $m_H$, and that subsequently
the meson--melting temperature is higher than for the conformal case.

\end{abstract}
\newpage \baselineskip=18pt \setcounter{footnote}{0}

\section{Introduction}
%

In making further progress in strengthening the connection between the
string theory based holographic techniques and experimentally
accessible strongly coupled systems such as the quark--gluon plasma
studied in heavy ion collisions at facilities such as RHIC and LHC
(for reviews see {\it e.g.,} refs.~\cite{Johnson:2010zzb,Jacak:2010zz}
and references therein), better understanding is needed of systems
well away from the conformal $\mathcal{N}=4$ theory. While it is
remarkable that finite temperature $\mathcal{N}=4$ theory (at large
$N_c$, for $SU(N_c)$ gauge group) already has several properties in
common with the quark gluon plasma ({\it e.g.,} the strikingly small
shear viscosity to entropy density ratio for its hydrodynamic
behaviour\cite{Policastro:2001yc,Kovtun:2004de}), the introduction of
temperature, while indeed breaking conformal symmetry and
supersymmetry, introduces only one scale into the problem, in terms of
which all other thermal properties are determined. This is evident in
the dual geometry, which is AdS$_5$--Schwarzschild (times $ S^5$)
\cite{Witten:1998qj,Witten:1998zw}, with an horizon at a radius that
sets the temperature. That system (and its dual geometry --- in local
AdS coordinates) has no deconfining phase transition at some non--zero
$T_c$, and is in its ``high'' temperature phase for any $T>0$.

In contrast, the microscopic theory of nuclear matter, quantum
chromodynamics (QCD), dynamically generates a natural scale,
$\Lambda_{\rm QCD}$, as a result of asymptotic freedom.  $\Lambda_{\rm
  QCD}$, ultimately helps determine the basic size of the bound states
of quarks and gluons in the theory, setting a finite transition
temperature for deconfinement to the quark--gluon plasma (QGP)
phase. This extends to mesons in the theory, even after the QGP has
formed. The properties of the spectrum of ``quarkonium'', as a
function of quark mass, and the subsequent melting of the mesons at
temperatures $T_{\rm melt}>T_c$, is a subject of considerable
interest, both experimentally and theoretically, not the least because
quarkonium suppression could be a valuable diagnostic probe in QGP
studies, were it sufficiently computationally understood.

While there are many active discussions of the extraction of $T_{\rm
  melt}$ in the literature (see {\it e.g.,}
ref.\cite{Grigoryan:2010pj}'s review section and references therein)
the analogous quantity in (so--called ``top--down'') holographic duals
is easy to extract (at large~$N_c$, and in a quenched limit where the
quarks do not back--react on the physics, so things are much
simpler). There, meson melting is a first order phase transition
\cite{Babington:2003vm,Albash:2006ew,Mateos:2006nu} from D7--brane
probe\cite{Karch:2002sh} fluctuations with boundary conditions leading
to infinitely long--lived bound states (the mesons) to fluctuations of
D7--branes that end on a black hole, where there the boundary
condition gives states that decay by falling into the black hole (the
``quasinormal'' modes --- the melted mesons). Nevertheless, the
placing of D7--branes in the AdS$_5$--Schwarzschild background (in the
zero back--reaction limit) will still result in a melting temperature
that recalls too much of the conformal nature of the parent
theory. The $\beta$--function of the resulting $\mathcal{N}=2$ theory
at zero quark mass is again vanishing (since it is of order $N_f/N_c$
and here $N_c>> N_f$ in the probe/quenched limit) and so $T_{\rm
  melt}$ will be determined only in terms of the quark masses, and
since there is no intrinsic non--thermal scale like $\Lambda_{\rm
  QCD}$, will be lower than what would be expected experimentally.

To make progress it is highly desirable to move away from the
neighbourhood of a conformal system and cleanly study meson melting in
a holographic dual of a gauge theory with a natural scale present
already at $T=0$. The mesons size and binding should depend upon this
scale (in addition to the bare quark masses) and so the gauge theory
would have a $T_{\rm melt}$ that is higher than if it had no scale.

This is the journey begun in our previous paper\cite{Albash:2011nw},
and continued in the present one. The gauge theory in question is
$\mathcal{N}=2^\ast$, obtained by making massive an $\mathcal{N}=2$
hypermultiplet within the $\mathcal{N}=4$ vector multiplet, and then
flowing (already starting at strong coupling) to the infra--red (IR)
to study the low energy physics. Now, there is no finite $\Lambda_{\rm
  QCD}$ in the IR for this theory, and instead the natural scale is
set by the hypermultiplet mass $m_H$, appearing in the dual geometry
as a structure that is an example of the enhan\c con
mechanism\cite{Johnson:1999qt}, a beautiful combination of
large~$N_c$, quantum effects, and strong coupling. The natural
coupling, $\lambda=g_{\rm YM}^2 N_c$, already starts out large for
this theory (in order to be accessible as a smooth gravity dual at
least asymptotically) and so dimensional transmutation generates no
natural scale $\Lambda_{\rm QCD}$ in the IR. This is the scale that
normally generates (through quantum effects) the separation between
monopole/dyon points on the Coulomb branch \emph{\`a la}
Seiberg--Witten\cite{Seiberg:1994aj,Seiberg:1994rs}. At large $N_c$,
there generically would be $N_c$ such points and since $\Lambda_{\rm
  QCD}$ is vanishing here they are densely packed together, spreading out to
form a circle, or in the natural coordinates, a line segment (see
refs.\cite{Evans:2000ct,Buchel:2000cn}). The radius of the circle (or
length of the line) is set by $m_H$, so the hypermultiplet mass $m_H$
plays the role of $\Lambda_{\rm QCD}$ for us in the
$\mathcal{N}=2^\ast$ theory, where the dual geometry is seen to be describing
part of the Coulomb branch of the theory. In the dual geometry (known
as the Pilch--Warner geometry\cite{Pilch:2000ue}), the Coulomb branch
is the place where the potential of a probe D3--brane vanishes, and
the enhan\c con is the large $N_c$ locus where their tension drops to
zero\cite{Evans:2000ct,Buchel:2000cn}. More generally,  in the
natural supergravity coordinates, the hypermultiplet mass $m_H$ sets a
radius in the core of the ten--dimensional geometry where the metric
becomes singular as one of the fields, $\chi$, which in the
ultra--violet (UV) sets the mass of the hypermultiplet, diverges.

In our previous paper\cite{Albash:2011nw} we holographically studied
quarks and mesons in the $\mathcal{N}=2^\ast$ theory at $T=0$ by
exploring D7--brane embeddings in the dual ten--dimensional
Pilch--Warner geometry. As we shall discuss later, our results already
contained the seeds of the anticipated finite temperature result --- a
higher $T_{\rm melt}$ than for the conformal theory --- and our
explicit study of finite temperature embeddings in this paper confirm
this, also uncovering a great deal of interesting and useful
detail. We use the finite temperature geometry generalizing the
Pilch--Warner geometry that was found in ref.\cite{Buchel:2003ah}, and
which has been studied extensively in the context of models of
QGP--like dynamics in
refs.~\cite{Buchel:2003ah,Buchel:2004hw,Buchel:2006bv,Buchel:2007vy,Buchel:2008uu}).
Our explorations allow us to study a family of D7--brane embeddings
and extract from them the behaviour of the meson melting temperature
$T_{\rm melt}$ as we vary quark mass and $m_H$.

In section \ref{sec:T=0}, we briefly review the zero temperature
background to establish conventions and notation.  In section
\ref{sec:T!=0}, we review the thermal version of the
$\mathcal{N}=2^\ast$ background.  In section \ref{sec:D7}, we
probe the background with D7--branes, extracting in
subsection~\ref{sec:condense} the condensate of the $\Delta = 3$
operator given by \cite{Kobayashi:2006sb}:
\begin{equation}
\langle \mathcal{O} \rangle = i \tilde{\psi} \psi + \tilde{q} \left( m_q + \sqrt{2} \phi_3 \right) \tilde{q}^\dagger + q^\dagger \left( m_q + \sqrt{2} \phi_3 \right) q + \mathrm{h.c.} \ ,
\end{equation}
where $\phi_3$ is the complex scalar in the $\mathcal{N} = 2$ vector
multiplet and the bare mass $m_q$ is the source of the operator. The
fields $\phi$ and $(q,\tilde{q})$ are the fermionic and bosonic
components of the quark multiplet. We discuss meson spectrum and the
meson--melting temperature in subsection~\ref{sec:melting}, and
conclude in section~\ref{sec:conclusions}.
%
\section{Zero Temperature ${\cal N}=2^\ast $ and Dual Extremal Geometry} \label{sec:T=0}
%
The matter content of the $\mathcal{N}=4$ supersymmetric Yang--Mills
theory consists of a gauge multiplet containing the bosonic fields
$(A_\mu, X_j)$, $j=1,\dots, 6$, where $X_j$ are real scalars
transforming as the $\mathbf{6}$ of $SO(6)$, and the fermionic fields
$(\lambda_j)$, $j = 1, \dots, 4$, which transform as the $\mathbf{4}$
of $SU(4)$.  Writing this matter content in terms of $\mathcal{N}=1$
superfields, the theory has one vector supermultiplet $(A_\mu,
\lambda_4)$ and three chiral multiplets $\Phi_j = (\lambda_j, \phi_j =
X_{2j - 1} + i X_{2 j})$, $j = 1, \dots, 3$. We can make two of the
chiral multiplets massive ($\phi_1$ and $\phi_2$, say, with equal
mass), preserving only an ${\cal N}=2$.

In the dual gravity picture, this will correspond to turning on two
real scalars $\alpha$ and $\chi$ with conformal dimension
$\Delta_\alpha =2$ and $\Delta_{\chi}=3$ respectively. The theory
described by the resulting flow to the infrared (IR) is called the
${\cal N}=2^\ast$ theory. The flow can be holographically described in
terms of five dimensional gauged supergravity, with action given by:
\begin{equation}
S_{5d} = \frac{1}{16 \pi G_{5}} \int d^5 x \sqrt{-G} \left( \mathcal{R} - 12 \left( \partial \alpha \right)^2 - 4 \left( \partial \chi \right)^2 - 4 \mathcal{P} \right) \ ,
\end{equation}
where the potential $\mathcal{P}$ for the scalars $\chi$ and $\alpha$ is:
\begin{equation}
\mathcal{P} = \frac{g^2}{16} \left[ \frac{1}{3} \left( \frac{ \partial W}{\partial \alpha} \right)^2 + \left( \frac{\partial W}{\partial \chi} \right)^2 \right] - \frac{g^2}{3} W^2 \ ,\quad {\rm with}\quad  W = - e^{- 2 \alpha} - \frac{1}{2} e^{4 \alpha} \cosh\left( 2 \chi \right)  \ .
\end{equation}
The constant $g$ is related to the AdS radius $R$, $g^2 = 4 / R^2$.

In the zero temperature (supersymmetric) case, the problem of solving
for  $\chi(r)$, $\rho(r)$ and $A(r)$ where the latter is
given by the (Einstein frame) metric ansatz:
\begin{equation}
ds_5^2 = e^{2 A} \left( - dt^2 + d \vec{x}^2 \right) + dr^2 \ ,
\end{equation}
reduces to solving the following first order equations \cite{Pilch:2000ue}:
\begin{eqnarray}
\frac{d \rho}{d r} &=& \frac{\rho}{3 R} \left( \frac{1}{\rho^2} - \rho^4 \cosh( 2 \chi ) \right) \ , \\
\frac{d A}{d r} &=& \frac{2}{3 R} \left( \frac{1}{\rho^2} + \frac{1}{2} \rho^4 \cosh( 2 \chi ) \right) \ , \\
\frac{d \chi}{d r} &=& -\frac{1}{2 R}  \rho^4 \sinh( 2 \chi ) \ .
\end{eqnarray}
Partial solutions are given by \cite{Pilch:2000ue}:
\begin{equation}
e^A = \frac{ k \rho^2}{\sinh(2 \chi)} \ , \quad \rho^6 = \cosh(2 \chi) + \sinh^2 (2 \chi) \left( \ln \left( \tanh(\chi) \right) + \gamma \right) \ .
\end{equation}
Here $k = m_H R$, where $m_H$ is the mass of the hypermultiplet and $\gamma$ is a constant corresponding to different slices through the moduli space of $\mathcal{N}=2^\ast$ in the IR.

Note that we can have a complete analytical solution for the metric in
terms of the field $\chi$ by making the following transformation:
\begin{equation}
d \chi = \frac{d \chi}{dr} d r  \rightarrow  dr^2 = \frac{4 R^2}{\rho^8 \sinh^2(2 \chi)} d \chi^2 \ ,
\end{equation}
such that the AdS boundary is at $\chi = 0$. In the interior $\chi \to
\infty$, and the apparently singular physics there was made sense of
in terms of the Coulomb branch of the gauge theory using D3--brane
probe techniques\cite{Buchel:2000cn,Evans:2000ct}. (For example, the
enhan\c con locus appears there, as mentioned in the introduction.)

Beyond that, we can solve for $\chi$ numerically, and it is convenient to do this
in a coordinate system given by:
\begin{equation} \label{eqt:z_hat}
\hat{z} = \frac{z}{R} = e^{-r / R} \ .
\end{equation}
It was shown in ref.~\cite{Buchel:2003ah} that in terms of this
coordinate, the fields have an expansion near the AdS boundary given
by (for $\gamma = 0$):
\begin{eqnarray}
\chi &=& k \hat{z} \left[ 1 + k^2 \hat{z}^2 \left( \frac{1}{3} + \frac{4}{3} \ln (k \hat{z}) \right) + k^4 \hat{z}^4 \left(- \frac{7}{90} + \frac{10}{3} \ln (k \hat{z}) + \frac{20}{9} \ln(k \hat{z})^2 \right) + \mathcal{O} (k^6 \hat{z}^6 \ln(k \hat{z})^3) \right] \ , \nonumber \\
\rho &=& 1 + k^2 \hat{z}^2 \left( \frac{1}{3} + \frac{2}{3} \ln( k \hat{z}) \right) + k^4 \hat{z}^4 \left( \frac{1}{18} + 2 \ln (k \hat{z}) + \frac{2}{3} \ln(k \hat{z})^2 \right) + \mathcal{O} ( k^6 \hat{z}^6 \ln( k \hat{z})^3 ) \ , \nonumber \\
A &=& - \ln (2 \hat{z}) - \frac{1}{3} k^2 \hat{z}^2 - k^4 \hat{z}^4 \left( \frac{2}{9} + \frac{10}{9} \ln(k \hat{z}) + \frac{4}{9} \ln( k \hat{z})^2 \right) + \mathcal{O} (k^6 \hat{z}^6 \ln( k \hat{z})^3) \ .
\end{eqnarray}

\section{Non--extremal Background} \label{sec:T!=0}
%
We follow the analysis of ref.~\cite{Buchel:2007vy}, which we repeat
here for completeness. Consider this form for the metric:
\begin{equation}
ds_5^2 = - c_1(r)^2 dt^2 + c_2(r)^2 d \vec{x}^2 + dr^2 \ .
\end{equation}
With this ansatz the equations of motion for the scalars give, in
terms of the scalar potential $\mathcal{P}$:
\begin{eqnarray}
\alpha''(r) + \partial_r \left( \ln \left( c_1(r) c_2(r)^3 \right) \right) \alpha'(r) - \frac{1}{6} \frac{\partial \mathcal{P}}{\partial \alpha} = 0 \ ,
\nonumber \\
\chi''(r) + \partial_r \left( \ln \left( c_1(r) c_2(r)^3 \right) \right) \chi'(r) - \frac{1}{2} \frac{\partial \mathcal{P}}{\partial \chi} = 0 \ ,
\end{eqnarray}
and the Einstein equations give:
\begin{eqnarray}
&& c_1''(r) + \partial_r \left( 3 \ln c_2 \right) c_1'(r) + \frac{4}{3} c_1(r) \mathcal{P} = 0 \ ,
\nonumber \\
&& c_2''(r) + \partial_r \left( \ln \left( c_1 c_2^2 \right) \right) c_1'(r) + \frac{4}{3} c_2(r) \mathcal{P} = 0 \ ,
\nonumber \\ \label{eqt:constraint}
{\rm and}\quad &&\alpha ' (r)^2 + \frac{1}{3} \chi ' (r)^2 - \frac{1}{3} \mathcal{P} - \frac{1}{2} \partial_r \left( \ln c_2(r) \right) \partial_r \left( \ln c_1(r) c_2(r) \right) = 0 \ ,
\end{eqnarray}
where for the last equation we used the previous two equations to simplify its form.
Since we will be solving these equations numerically, the AdS radial
coordinate $r$ turns out not to be convenient, since it runs to
infinity (the location of the AdS boundary).  Therefore, we define a
coordinate $x(r)$ satisfying:
\begin{equation}
1 - x(r) = \frac{c_1(r)}{c_2(r)} \ ,
\end{equation}
such that $x \in \left[ 0 , 1 \right]$, with $x = 0$ being the AdS
boundary and $x = 1$ being the event horizon.  An important thing to
note is that this change of coordinates will only work for the
non--extremal case, since for the extremal case, we have $c_1(r) =
c_2(r)$ and so $x = 0$ for all $r$.  Using equation
\reef{eqt:constraint}, we can derive an expression for $ d x / d r$
entirely in terms of $x$:
\begin{equation}
\frac{ d x } {d r } = \pm \sqrt{ \frac{ 2 \mathcal{P} c_2(x)^2 \left( 1 - x \right) } {\left(1-x \right) c_2(x)^2 \left( 6 \alpha'(x)^2 + 2 \chi'(x)^2 \right) - 3 c_2'(x) \left(- c_2(x) + 2 (1-x) c_2'(x) \right)}} \ .
\end{equation}
In this new coordinate system, the equations of motion reduce to:
\begin{eqnarray}
&& c_2''(x) - 5 \frac{c_2'(x)^2}{c_2(x)} + \frac{c_2'(x)}{1-x} + \frac{4}{3} c_2(x) \left( 3 \alpha'(x)^2 + \chi'(x)^2 \right) = 0 \ ,
\nonumber\\
&&\alpha''(x) - \frac{\alpha'(x)}{1-x} - \frac{1}{6} G_{xx} \frac{\partial P}{\partial \alpha} = 0 \ ,
\nonumber \\
{\rm and}\quad &&\chi''(x) - \frac{\chi'(x)}{1-x} - \frac{1}{2} G_{xx} \frac{\partial P}{\partial \chi} = 0 \ ,
\end{eqnarray}
where
\begin{equation}
G_{xx} =   \frac{\left(1-x \right) c_2(x)^2 \left( 6 \alpha'(x)^2 + 2 \chi'(x)^2 \right) - 3 c_2'(x) \left(- c_2(x) + 2 (1-x) c_2'(x) \right)} { 2 \mathcal{P} c_2(x)^2 \left( 1 - x \right) }  \ .
\end{equation}
Note that in order to derive these results, one needs the quantity:
\begin{equation}
\partial_x G_{xx} = - \frac{4 c_2'(x) \left( -2(1-x) c_2(x)^2 \left( 3 \alpha'(x)^2 + \chi'(x)^2 \right) - 3 c_2(x) c_2'(x) + 6 (1-x) c_2'(x)^2 \right) }{\mathcal{P} (1-x) c_2(x)^3} \ ,
\end{equation}
where we have used the equations of motion to simplify the expression
in terms of only first derivatives of the fields. It is worth
recalling here the connection to the fields of the previous section:
\begin{equation}
\rho(x) = e^{\alpha(x)} \ , \quad c_2(x) = e^{A(x)} \ .
\end{equation}
%
\subsection{Non--extremal AdS$_5$}
Let us consider some special cases to orient ourselves.  Consider the
case of $\alpha = \chi = 0$.  The most general solution for $A(x)$ is:
\begin{equation}
A(x) = \ln a_1 - \frac{1}{4} \ln (x (2-x) + a_2) \ .
\end{equation}
We can take $a_2 = 0$, and the metric is now given by\footnote{There
  appears to be a typo in the $g_{xx}$ term for this equation in
  ref.~\cite{Buchel:2007vy}.}:
\begin{equation}
ds_5^2 = a_1^2 \left( x (2-x) \right)^{-1/2} \left( - \left( 1- x \right)^2 dt^2 + d \vec{x}^2 \right) + \frac{L^2}{4} \left( 2  x - x^2 \right)^{-2} dx^2 \ .
\end{equation}
We can return this to the perhaps more familiar form of AdS$_5$ Schwarzschild by
the following coordinate transformation:
\begin{equation}
x = 1 - \sqrt{ 1 - \frac{b^4}{u^4} } \ ,
\end{equation}
and the choice of $a_1 = b/R = \left( \pi T R \right) $ such that we have:
\begin{equation}
ds_5^2 = - \left( \frac{u^2}{R^2} - \frac{b^4}{u^2 R^2} \right) dt^2 + \frac{u^2}{R^2} d \vec{x}^2 + \left( \frac{u^2}{R^2} - \frac{b^4}{u^2 R^2} \right)^{-1} du^2 \ .
\end{equation}
%
\subsection{Non--extremal Pilch--Warner Geometry}
\begin{figure}[ht]
\begin{center}
\subfigure[SUSY Deformations]{\includegraphics[width=2.5in]{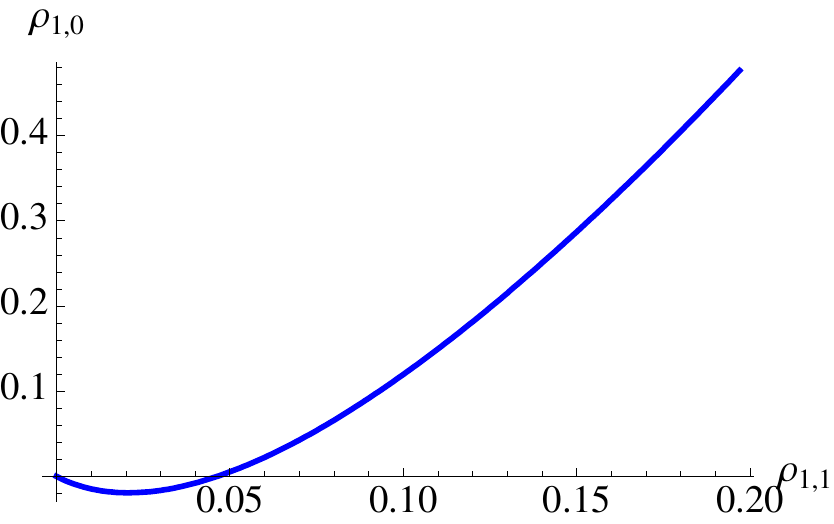}\label{fig:NEPW_SUSY1}} \hspace{0.5cm}
\subfigure[SUSY Deformations]{\includegraphics[width=2.5in]{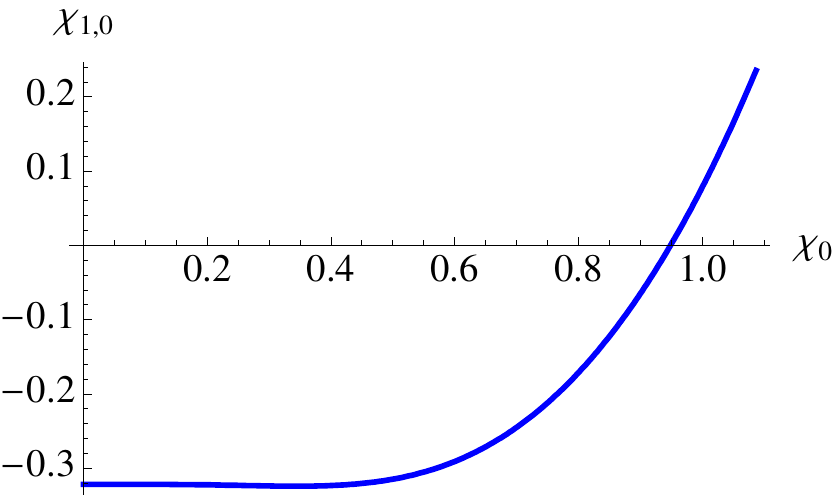}\label{fig:NEPW_SUSY2}}
   \caption{\small Non--Extremal Pilch--Warner solutions.}  \label{fig:NonExtremalPW}
   \end{center}
\end{figure}

%
Motivated by our result above for  non--extremal AdS$_5$, we consider an ansatz for the fields of the following form:
\begin{equation}
A(x) = \ln(\kappa) - \frac{1}{4} \ln(x(2-x)) + a(x) \ .
\end{equation}
Near the AdS boundary, the fields have the following leading behavior:
\begin{eqnarray}
\rho(x) &=& 1 + x^{1/2} \left( \rho_{10} + \rho_{11} \ln(x) \right) + O(x) \ , \nonumber \\
\chi(x) &=&  \chi_0 x^{1/4} \left( 1 + x^{1/2} \left( \chi_{10} + \frac{1}{3} \chi_0^2 \ln(x) \right) + O(x) \right) \ ,\nonumber \\
a(x) &=& - \frac{1}{9} \chi_0^2 x^{1/2} + O(x) \ .
\end{eqnarray}
The parameters $(\rho_{1,1}, \chi_0)$ are related to the bosonic and
fermionic masses ($m_B$ and $m_F$ respectively) of the components of the (would--be)
${\cal N}=2$ hypermultiplet) \cite{Buchel:2007vy}:
\begin{equation}
\left( \frac{m_B}{T} \right)^2 = \frac{24 \pi^2}{\sqrt{2}} e^{6 a_h} \rho_{1,1} \ , \quad \frac{m_F}{T} = 2^{3/4} \pi e^{3 a_h} \chi_0 \ ,
\end{equation}
where we have defined $a_h \equiv a(1)$.  Furthermore, the parameter
$\kappa$ is related to the temperature \emph{via}
\cite{Buchel:2007vy}:
\begin{equation} \label{eqt:temperature}
T = \frac{\kappa}{R \pi} e^{-3 a_h} \ .
\end{equation}
Therefore, to proceed, we fix the values of $(\kappa, \rho_{1,1},
\chi_0)$ and search the values of $(\rho_{1,0}, \chi_{1,0})$ that give
us regular solutions (\emph{i.e.} solutions for $(\rho(x), \chi(x),
a(x))$ that are have vanishing first derivative at the event horizon).
We focus our attention on a particular type of solution, which we
label as ref.~\cite{Buchel:2007vy} does as a ``supersymmetric
deformation'' with $m_F = m_B \equiv m_H$, which corresponds to taking:
\begin{equation}
\chi_0^2  = 6 \rho_{1,1} \ .
\end{equation}
We present our numerical solution for this case in figure
\ref{fig:NonExtremalPW}, and they appear to agree very well with the
results of ref.~\cite{Buchel:2007vy}.
We will not consider cases of the type $m_F\neq m_B$ in the rest of
this paper.
\section{Quarks and Mesons from D7--branes} \label{sec:D7}

The five dimensional physics may be oxidized to a ten dimensional
system (more suitable for discussion of the D--brane probing) within
the type~IIB supergravity \cite{Pilch:2000ue}, with the ten
dimensional metric (in Einstein frame) given by:
\begin{equation}
ds_{10}^2 = \Omega^2 d s_5^2+ \frac{a^2}{2} \frac{ \Omega^2}{\rho^2} \left(c^{-1} d\theta^2 + \rho^6 \cos^2 \theta \left(\frac{\sigma_1^2}{c X_2} + \frac{\sigma_2^2 + \sigma_3^2}{X_1} \right) + \sin^2 \theta \frac{d \phi^2}{X_2} \right) \ ,
\end{equation}
where:
\begin{equation} \begin{array}{rcl}
X_1(r, \theta) &=& \cos^2 \theta + \rho(r)^6 \cosh(2 \chi(r)) \sin^2 \theta \ , \\
X_2(r, \theta) &=& \cosh(2 \chi(r)) \cos^2 \theta + \rho(r)^6 \sin^2 \theta \ , \\
\Omega^2 &=& \frac{\left(c X_1 X_2\right)^{1/4}}{\rho} \ , \quad
a^2 = \frac{8}{g^2} = 2 R^2 \ , \\
c &=& \cosh(2 \chi) \ , \quad
\rho =  e^{\alpha} \ , 
\end{array}
\end{equation}
and there is a deformed $S^3$ with $SU(2)$ invariant 1--forms
\begin{equation}
\begin{array}{rcl}
  \sigma_1 &=& \frac{1}{2} \left( d\alpha + \cos \psi d \beta \right) \ , \\
  \sigma_2 &=& \frac{1}{2} \left( - \sin \alpha d \psi + \cos \alpha \sin \psi d \beta \right) \ , \\
  {\rm and}\quad \sigma_3 &=& \frac{1}{2} \left( \cos \alpha d \psi + \sin \alpha \sin \psi d \beta \right) \ ,
\end{array}
\end{equation}
and the dilaton is given by:
\begin{equation}
e^{-\Phi}  = \frac{\sqrt{c X_1 X_2}}{ c X_1 \sin^2 \phi + X_2 \cos^2 \phi} \ .
\end{equation}
There are several other fields, but we will not need them for the
embeddings we will choose, following our work at $T=0$ in
ref.\cite{Albash:2011nw}, to which we refer the reader for details.
Following our analysis in ref.~\cite{Albash:2011nw}, the relevant
action for us is:
\begin{equation} \label{eqt:action}
S_{D7} = - \mu_7 \int d^8 \xi \ e^{\Phi} \sqrt{- \det \left( P[G]_{ab} \right)} \ ,
\end{equation}
with a D7--brane embedding given by:
\begin{equation}
\xi^a = x^a \ , a = 0, \dots, 7 \ , \quad \theta \equiv \theta(x) \ , \quad \phi = \pi/2, 3 \pi/2 \ .
\end{equation}
Near the AdS boundary ($x \to 0$), the field $\theta(x)$ has asymptotic behavior given by:
\begin{equation} \label{eqt:theta_expansion}
\theta(x) = x^{1/4} \theta_0 + x^{3/4} \left( \theta_2 - \frac{1}{6} \chi_0^2 \theta_0 \ln x \right) + \dots \ .
\end{equation}
To extract the bare quark mass and the condensate, it is convenient to
express equation \reef{eqt:theta_expansion} in terms of the coordinate
$\hat{z}$ of equation \reef{eqt:z_hat}.  In order to do this, we
compare the expansion of $e^A$ near the AdS background for the zero
temperature background (which is in terms of $\hat{z}$) with that of
the finite temperature background (which is in terms of $x$), since
near the AdS boundary, the asymptotic behavior of both backgrounds
should be the same.  Using these two expansions and setting them
equal, we can solve (iteratively) for $x(\hat{z})$,
\begin{equation}
x( \hat{z} ) = 8 \kappa^4 \hat{z}^4 + \frac{128 \sqrt{2}}{9} \chi_0^2 \kappa^6 z^6 + \dots \ . 
\end{equation}
Plugging the expression for $x(\hat{z})$ into the asymptotic equation
of $\theta(x)$, we get:
\begin{equation}
\theta(x(\hat{z})) = \hat{z} \frac{k}{\chi_0} \theta_0 + \hat{z}^3 \left( \frac{k}{\chi_0} \right)^3 \left( \theta_2 + \frac{2}{9} \theta_0 \chi_0^2 - \frac{2}{3} \theta_0 \chi_0^2 \ln\left(\frac{k}{\chi_0} \right) - \frac{2}{3} \theta_0 \chi_0^2 \ln(\hat{z}) \right) + \dots \ .
\end{equation}
If we make the identification:
\begin{equation}
\hat{\theta}_0 = \frac{k}{\chi_0} \theta_0 \ , \quad \hat{\theta}_2 =  \left( \frac{k}{\chi_0} \right)^3 \left( \theta_2 + \frac{2}{9} \theta_0 \chi_0^2 - \frac{2}{3} \theta_0 \chi_0^2 \ln\left(\frac{k}{\chi_0} \right) \right) \ ,
\end{equation}
and use the expression for the bare quark mass and the condensate \cite{Albash:2011nw}:
\begin{equation}
m_q = \frac{R}{2\pi \alpha'} \hat{\theta}_0 \ , \quad \langle \mathcal{\bar{\psi}\psi} \rangle = 2 \pi^2 R^5 \mu_7 \left( - 2 \hat{\theta}_2 + \frac{1}{3} \hat{\theta}_0^3 - \frac{4}{3} k^2 \hat{\theta}_0 \ln \hat{\theta}_0 \right) \equiv 2 \pi^2 R^5 \mu_7 \hat{C} \ ,
\end{equation}
we find that the bare quark mass and condensate are given by:
\begin{equation} \label{eqt:C}
m_q = \frac{R \kappa}{2 \pi \alpha'} 2^{3/4} \theta_0 \ , \quad \hat{C} =  \left( \frac{k}{\chi_0} \right)^3 \left( - 2 \theta_0 + \frac{1}{3} \theta_0^3 - \frac{4}{9} \theta_0 \chi_0^2 - \frac{4}{3} \theta_0 \chi_0^2 \ln (\theta_0) \right)  \equiv \kappa^3 C \ ,
\end{equation}
where we have used that $k / \chi_0 = 2^{3/4} \kappa$.
%
%
\subsection{Solving for the Condensate}\label{sec:condense}
\begin{figure}[ht]
\begin{center}
\subfigure[Raw Result]{\includegraphics[width=2.5in]{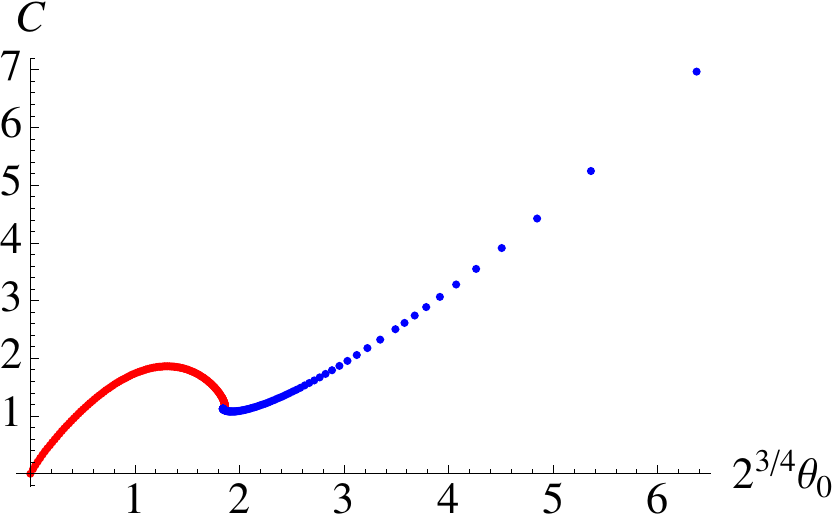}\label{fig:raw_result}} \hspace{0.5cm}
\subfigure[Result after $T=0$ subtraction]{\includegraphics[width=2.5in]{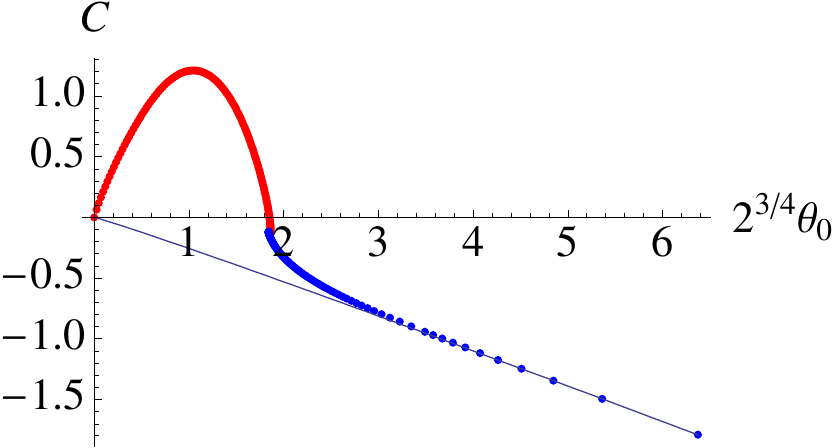}\label{fig:T=0_subtraction_result}} \hspace{0.5cm}
\subfigure[Physical Result]{\includegraphics[width=2.5in]{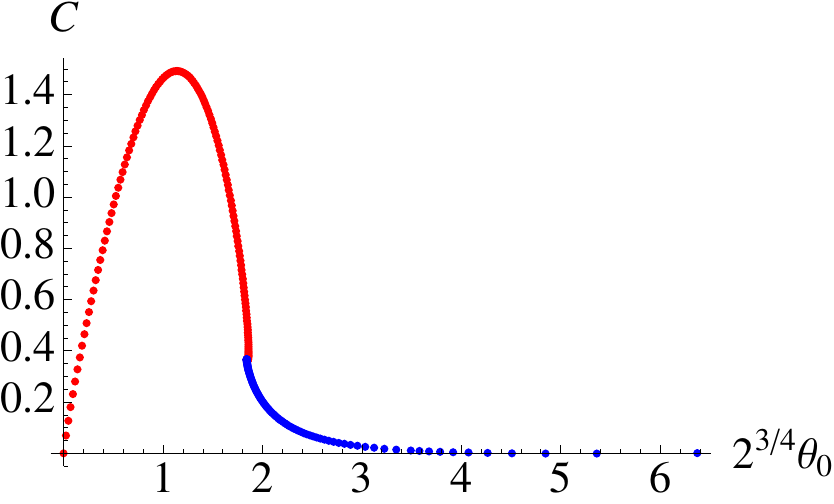}\label{fig:physical_result}} \hspace{0.5cm}
\subfigure[Several Condensate Curves]{\includegraphics[width=2.5in]{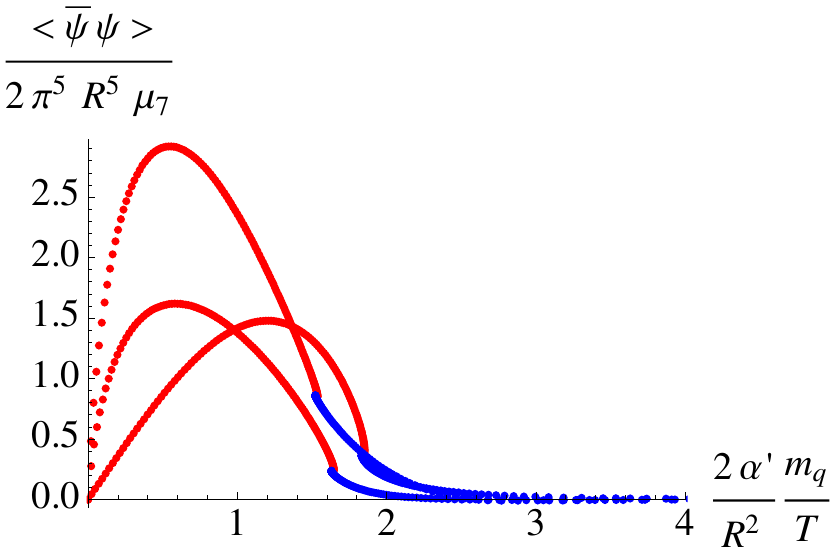}\label{fig:c_vs_m_physical}}
\caption{\small An example of the results (for $\frac{m_H}{T}=
  0.3193$) and subsequent subtraction scheme results.  In (a), we have
  our raw results.  In (b), the solid line is a curve of the form $a x
  + b x \ln x$ that is fit to the condensate for large bare quark.
  After subtracting this curve, we end up with the result in (c).  In
  (d), we give curves for $\frac{m_H}{T} = 0 , 5.12418, 6.93171$ in
  ascending order. Throughout, the black hole embeddings extend from
  the left, starting at the origin, while the Minkowski embeddings
  extend to the right. They meet at a kink that hides a region of
  multivaluedness discussed later in the
  text.}  \label{fig:NumericalResults}
   \end{center}
\end{figure}

%
We show the results for our condensate in figure \ref{fig:raw_result}.
We find that there is a divergence for large quark mass, a situation
that we have encountered in the zero temperature
case~\cite{Albash:2011nw}.  In the zero temperature case, the
divergence can be attributed to the numerical artifacts arising from
solving for $\theta$ on top of a numerical solution for $\chi$.  We
can perform a similar subtraction as in the zero temperature case to
get the result in figure \ref{fig:T=0_subtraction_result}.  We see
that there is still a divergence that is unaccounted for.  This is
unsurprising since for finite temperature, we have more numerics in
the background ($\rho$ and $a$).  In particular, the asymptotic
behavior of $a(x)$ is what gives the coefficient of both the linear
term and linear--log term in equation \reef{eqt:C}.  Therefore, we
should expect that numerical artifacts would appear as exactly linear
and linear--log terms.  This is indeed what we see in figure
\ref{fig:T=0_subtraction_result}.  We can subtract off this divergence
by fitting the asymptotic behavior, and get a physical result as shown
in figure \ref{fig:physical_result}.  We give plots in terms of
physical quantities in figure \ref{fig:c_vs_m_physical}.
%
%
\subsection{Meson Melting Temperature}\label{sec:melting}
\begin{figure}[ht!]
\begin{center}
\subfigure[Meson mass for fixed bare quark mass.]{\includegraphics[width=2.5in]{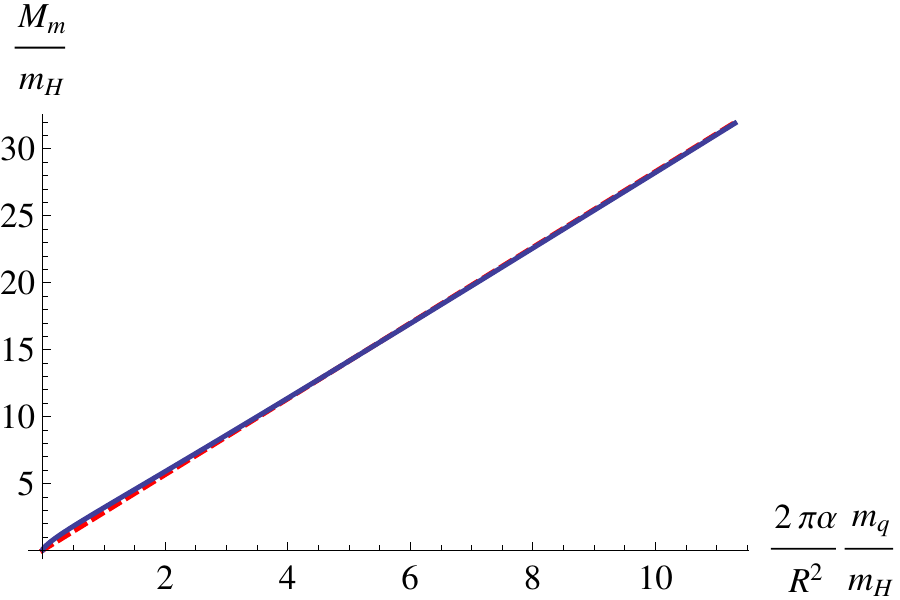}\label{fig:meson_T=0_mq}} \hspace{0.5cm}
\subfigure[Meson mass for fixed constituent mass.]{\includegraphics[width=2.5in]{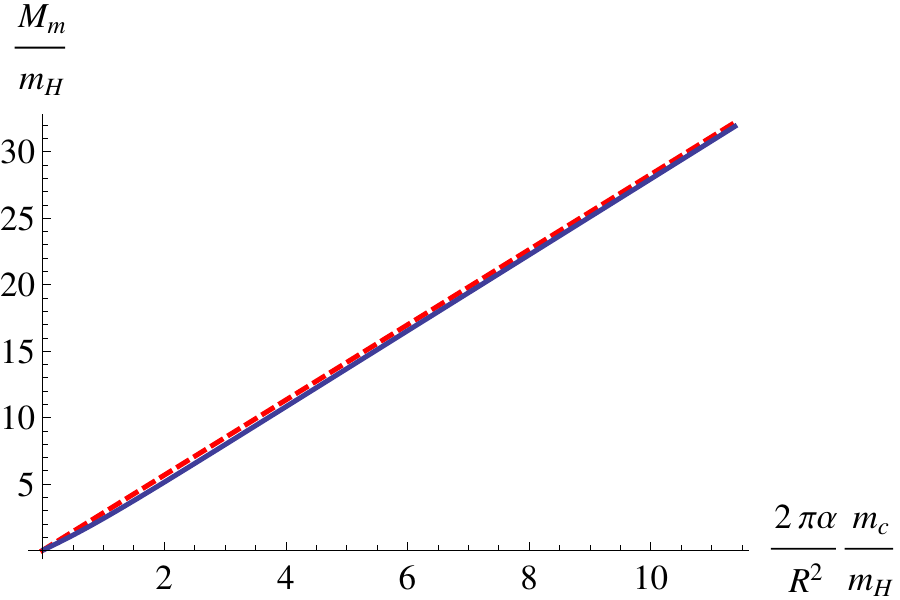}\label{fig:meson_T=0_mc}}
\subfigure[Detail of part (a).]{\includegraphics[width=2.5in]{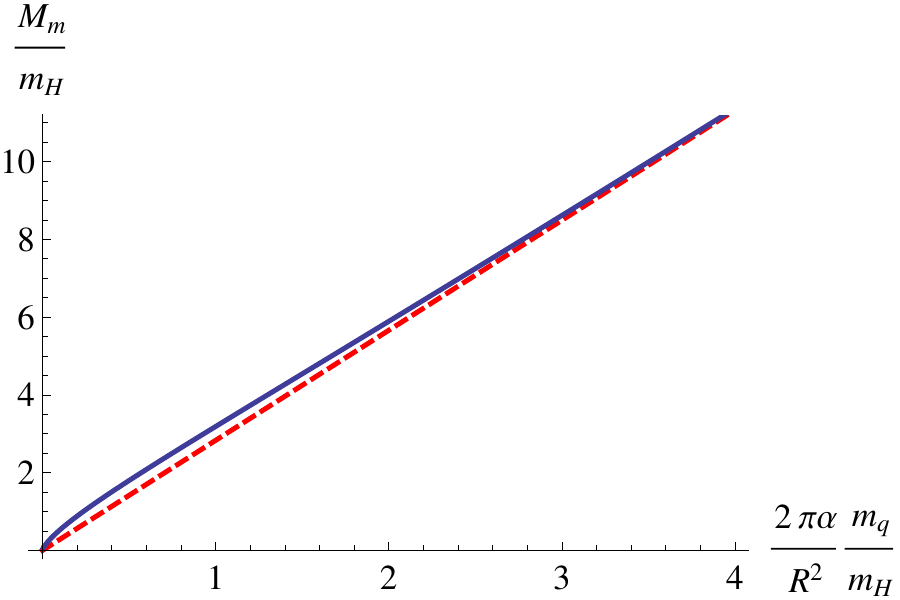}\label{fig:meson_T=0_mq_zoom}} \hspace{0.5cm}
\subfigure[Detail of part (b).]{\includegraphics[width=2.5in]{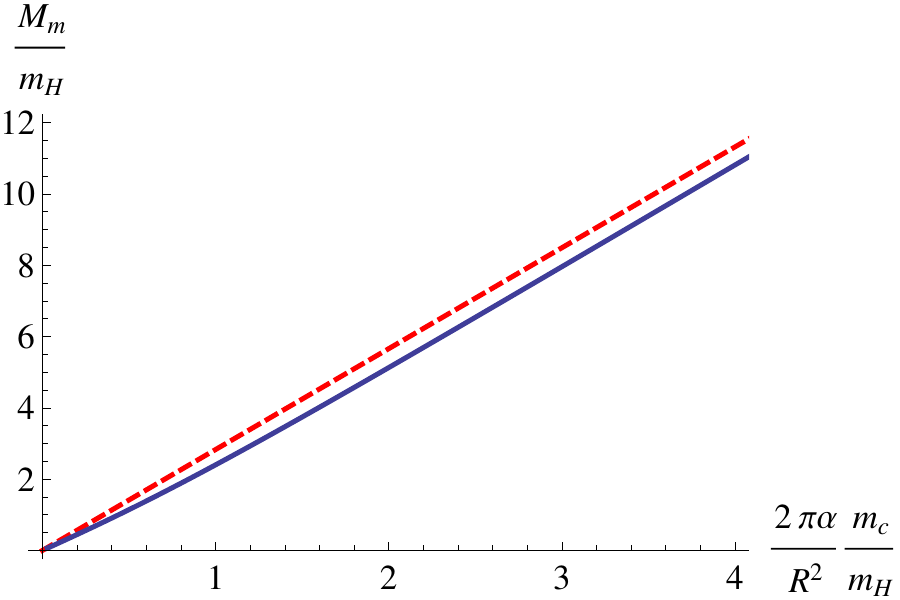}\label{fig:meson_T=0_mc_zoom}}
   \caption{\small Meson spectrum in the extremal $\mathcal{N}=2^\ast$ background.  The red, dashed line is the meson spectrum in the $\mathcal{N}=4$ background, whereas the blue, solid line is the meson spectrum in the $\mathcal{N}=2^\ast$ background.}     \label{fig:meson_T=0}   \end{center}
\end{figure}

%
We now wish to study the effect of the hypermultiplet mass on the
melting temperature of the mesons, as discussed in the
introduction. The meson is associated with 7--7 strings and its mass
can be determined by considering fluctuations of the D7--brane
embedding \cite{Karch:2002xe}.  Recall that the Minkowski embeddings
are related to the phase where we have mesons (quark bound states) in
our gluon plasma, whereas the black hole embeddings are related to the
phase where the mesons have melted in the plasma (the finite lifetime
quasinormal modes arising from in--falling boundary conditions at the
horizon).  We may predict what might happen based on revisiting the
zero temperature result.  We revisit our results from
ref.~\cite{Albash:2011nw}, but present them in a different fashion in
figure \ref{fig:meson_T=0}.

Recall that the meson mass should in general be given by:
\begin{equation}
M_m = 2 m_c - E_b \ ,
\end{equation}
where $m_c$ is the constituent quark mass and $E_b$ is the (positive) binding energy.  The constituent mass is calculated from the Nambu--Goto action per unit time for the string that hangs (in the $x$ direction) from the ``end'' of the D7--brane at $\theta(x = x_0) = \pi/2$ to the event horizon at $x=1$:
\begin{equation}
m_c = \frac{2 }{2 \pi \alpha'} \int^1_{x_0} d x e^{\Phi/2} \sqrt{-g_{tt} g_{xx}} \Big|_{\theta = \pi/2}
\end{equation}
In these scenarios, the meson mass is a factor of $\lambda^{-1/2}$
relative to the constituent mass, so the meson is said to be deeply
bound~\cite{Kruczenski:2003be}.  For this reason, we cannot directly
extract the binding energy for the meson at a given constituent mass,
however  we can make relative comparisons between the
constituent mass in $\mathcal{N}=2^\ast$ versus the constituent mass
in the $\mathcal{N}=4$ for fixed meson mass.  Our result in figure
\ref{fig:meson_T=0} suggests that the same mass meson in
$\mathcal{N}=2^\ast$ has a higher constituent mass than the meson in
$\mathcal{N}=4$.  This in turn suggests that the (positive) binding
energy in $\mathcal{N}=2^\ast$ is larger.  Therefore. we would predict
that in the thermal case, we should expect a raising in the melting
temperature.

\begin{figure}[ht]
\begin{center}
\subfigure[Meson spectrum for fixed bare quark mass]{\includegraphics[width=2.5in]{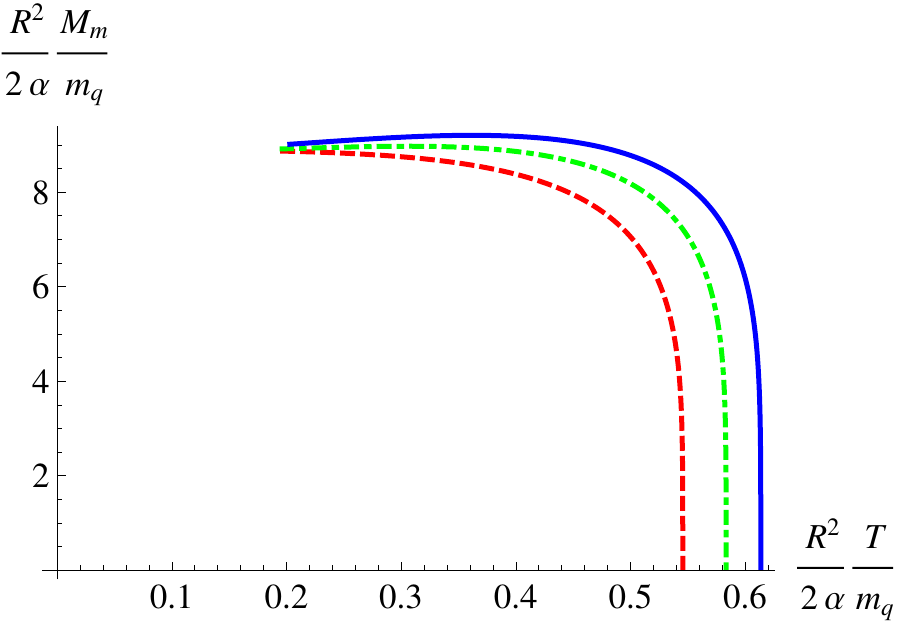}\label{fig:meson_spectrum_mq}} \hspace{0.5cm}
\subfigure[Meson spectrum for fixed constituent mass]{\includegraphics[width=2.5in]{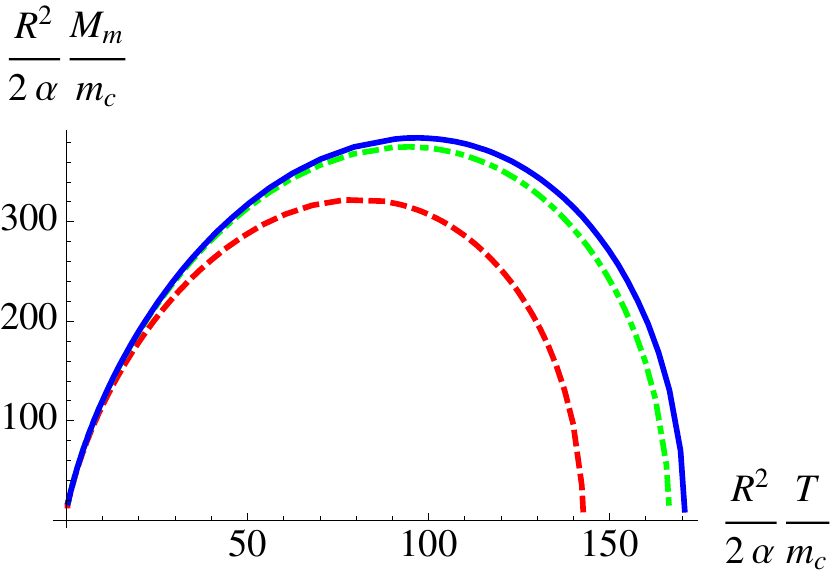}\label{fig:meson_spectrum_mc}}
   \caption{\small Solutions for the meson spectrum for different values of $m_H/T$.  The curves correspond to the following values of $m_H/T$: (Red dashed, 0), (Green dot--dashed, 3.74), (Blue solid, 5.12).}  \label{fig:meson_spectrum}
   \end{center}
\end{figure}

To proceed, we can study the meson mass associated with different
Minkowski embeddings as follows.  We consider fluctuations of the
field $\theta(x)$ such that:
\begin{equation}
\theta(x) = \theta_0(x) + 2 \pi \alpha' \Phi(t,x) \ ,
\end{equation}
where $\theta_0(x)$ corresponds to the embedding solution.  We can
expand the Dirac--Born--Infeld (DBI) action given previously (see equation
\reef{eqt:action}) to quadratic order in $\Phi(t,x)$ and derive the
equations of motion for $\Phi(t,x)$ from the resulting Lagrangian.  We
take as an ansatz for the time dependence of $\Phi(t,x)$:
\begin{equation}
\Phi(t,x) = e^{- i \frac{\kappa}{R} \tilde{\omega} t} \phi(x) \ ,
\end{equation}
such that the meson mass (associated to this fluctuation of the
D7--brane) is given by:
\begin{equation}
M = \frac{2 \kappa}{R} \tilde{\omega} \ .
\end{equation}
Note that the factor of 2 is because of the factor of 1/4 associated
with $G_{tt}$ in the UV.  The allowed values for $\tilde{\omega}$ are
computed by requiring that $\phi(x)$, which has expansion in the UV
given by:
\begin{equation}
\phi(x) \to x^{1/4} \phi_1 + x^{3/4} \phi_2 \ ,
\end{equation}
only has normalizable modes, \emph{i.e.} $\phi_1 = 0$.  We present
some solutions for the meson mass in figures~\ref{fig:meson_spectrum}.
Several important features include the crossing over of the meson mass
ratio $M_m / m_q$ below that of thermal $\mathcal{N}=4$ as we approach
$T/m_q=0$ in figure \ref{fig:meson_spectrum_mq}.  This was already
observed in the zero temperature results in
figure~\ref{fig:meson_T=0_mq}.

\begin{figure}[ht!]
\begin{center}
\includegraphics[width=2.5in]{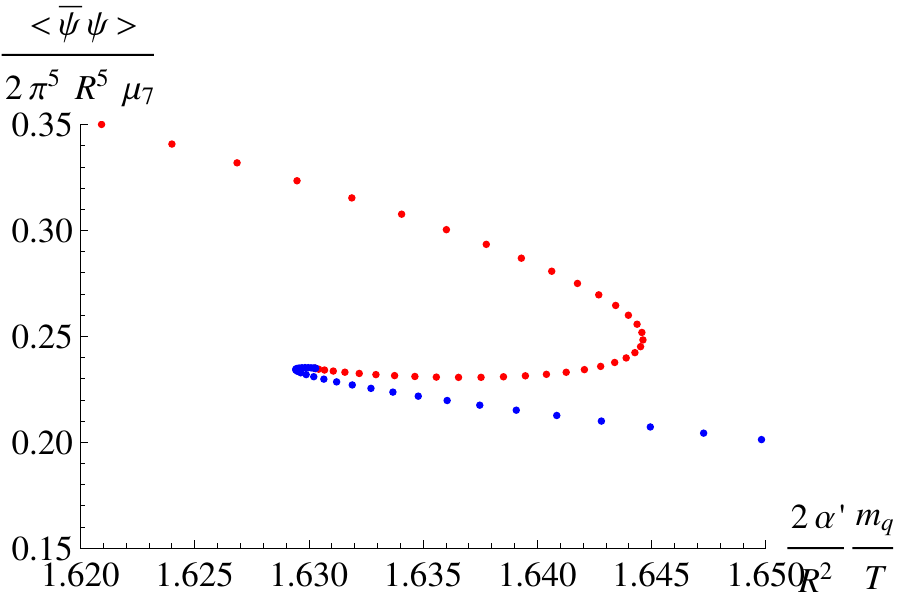}
\caption{\small An enlarged portion of a condensate {\it vs.} mass
  curve showing the multivaluedness in the region where the melting
  transition occurs. The black hole embeddings enter from the left,
  while the Minkowski embeddings enter from the
  right.}  \label{fig:zoom} \end{center}
\end{figure}

The meson melting transition occurs near where the meson mass is zero
(the far right of figures~\ref{fig:meson_spectrum}), but not exactly.
To find the exact location (in terms of the quark mass), we need to
calculate the free energy of the system and find where the jump
(determined by minimizing the free energy) between the black hole and
Minkowski embeddings occurs, inside the region of multivaluedness of
the condensate {\it vs.} mass curves of
figure~\ref{fig:c_vs_m_physical}. We have enlarged an example of such
a region in figure~\ref{fig:zoom}. (See {\it e.g.,}
ref.~\cite{Albash:2006ew} for examples of this extraction of the
temperature.)

\begin{figure}[ht]
\begin{center}
\subfigure[Melting temperature for fixed bare quark mass]{\includegraphics[width=2.5in]{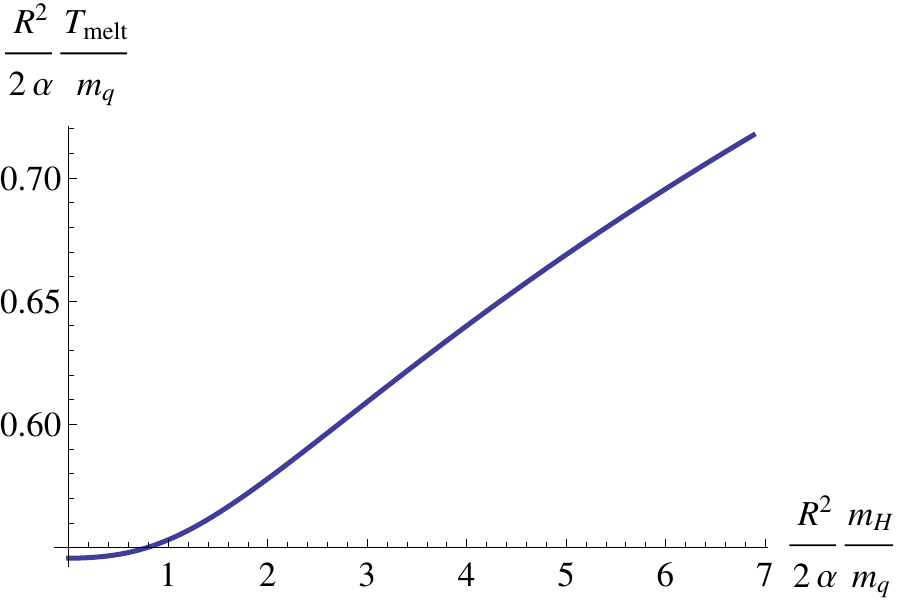}\label{fig:T_melt_mq}} \hspace{0.5cm}
\subfigure[Melting temperature for fixed constituent mass]{\includegraphics[width=2.5in]{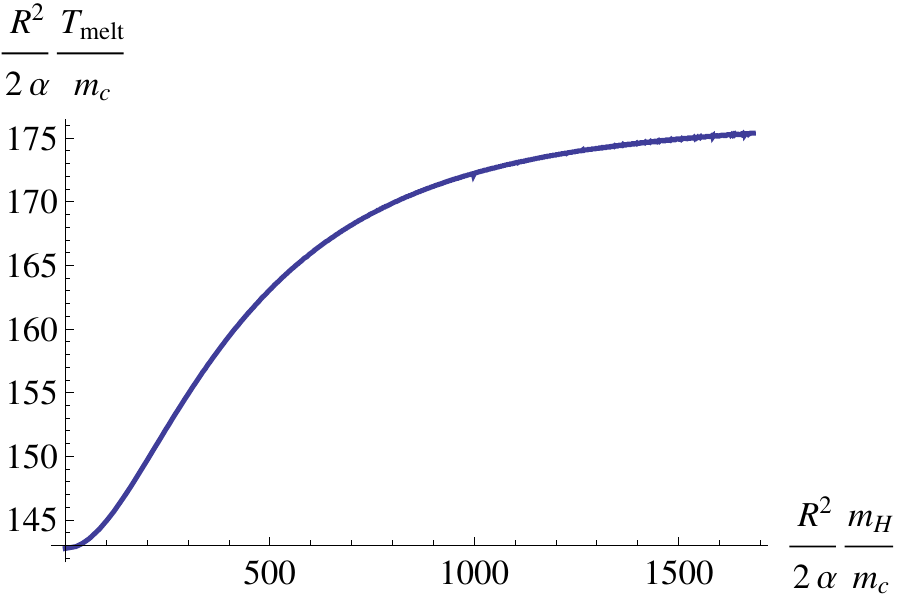}\label{fig:T_melt_mc}}
   \caption{\small The melting temperature for fixed bare and constituent mass.}  \label{fig:T_melt}
   \end{center}
\end{figure}

We do not have numerical control of the free energy. However, where
the meson mass drops to zero (corresponding to the end of the
Minkowski embeddings) provides a good estimate of the melting
temperature especially in terms of the bare quark mass.  This is
because the crossover region between the black hole and Minkowski
embeddings is a very small part of the overall curve (this becomes
even more accurate as $m_H$ increases).  In terms of the constituent
mass, we believe that the general behavior is correct although we
should not take the exact numerical values here too seriously since
the constituent mass rises very rapidly near the zero mass meson.
With this in mind, we can systematically find an estimate of the
melting temperature, and we present our results in figure
\ref{fig:T_melt}.
As predicted by the $T = 0$ results, the melting temperature increases
as we increase $m_H$.  Notice that the melting temperature saturates
for very large $m_H$, a fact that we discuss further in the conclusions.

\section{Conclusions} \label{sec:conclusions}

In a very clean setting, we have been able to demostrate how the move
to a non--conformal system changes the quark and meson dynamics in an
important way. The instrinsic scale, $m_H$, of the
$\mathcal{N}=2^\ast$ theory affects the temperature at which mesons of
a given mass melt (an important phenomenon in {\it e.g.,} the
experimentally accessible quark--gluon plasmas studied at RHIC and
LHC). This is because the quark bound states have their properties
determined by not just the scale set by the bare quark mass, but also
by $m_H$. There is a strong analogy to QCD, with its dynamically
generated scale $\Lambda_{\rm QCD}$, and indeed this was in part the
motivation for our study.

Of course, the $\mathcal{N}=2^\ast$ theory is still quite different
from QCD, and so we should expect differences in the meson
dynamics. For example, the scale of $\mathcal{N}=2^\ast$ is the mass
of a field in the theory, while in QCD the scale is generated by
dimensional transmutation.  Indeed, we see from
figure~\ref{fig:T_melt_mc} that, for fixed constituent quark mass,
$T_{\rm melt}$ eventually saturates for high $m_H$, a non--trivial
dependence that one would not expect in QCD. We expect that this is
due to the fact the the binding energy and constituent quark mass get
their contributions from the massive hypermultiplet in different
ways. The contribution to binding energy evidently becomes more
suppressed at higher~$m_H$. If this were a weakly coupled scenario, an
examination of the structure of the Feynman diagrams contributing to
each quantity at a given order (keeping only planar diagrams, since we
are at large $N_c$) might make this more manifest, but it is not clear
if the physics of interest is at all visible in perturbation theory,
since we are dealing with bound states. It would be interesting to
investigate this further.

Another key difference is that while we did have a scale, we were not
working in a confining theory. The geometry tells us about the Coulomb
branch of the $\mathcal{N}=2^\ast$ theory. To get access to quark and
meson dynamics (including meson melting) in a confining theory, a
study of how to holographically lift the $\mathcal{N}=2^\ast$ theory's
Coulomb branch and trigger monopole condensation would be one of a
number of interesting directions to pursue.

\section*{Acknowledgments}
This work was supported by the US Department of Energy and the USC
Dana and David Dornsife College of Letters, Arts, and Sciences. We
thank Hovhannes Grigoryan and Peter Steinberg for valuable comments.
%

%

\providecommand{\href}[2]{#2}\begingroup\raggedright\endgroup

\end{document}